%% file: beholder.tex
\documentclass[journal]{vgtc}                     

\onlineid{1260}
\vgtccategory{Research}
\vgtcpapertype{evaluation}

\title{Eye of the Beholder: Towards Measuring Visualization Complexity}

\author{%
  \authororcid{Johannes Ellemose}{0009-0006-9676-6204} and
  \authororcid{Niklas Elmqvist}{0000-0001-5805-5301}
}

\authorfooter{
  \item Johannes Ellemose and Niklas Elmqvist are with the Department of Computer Science at Aarhus University, Aarhus, Denmark.
        E-mail: \{ellemose, elm\}@cs.au.dk.
}

\abstract{%
  \input{sections/abstract}
}


\teaser{
  \centering
  \includegraphics[alt={An arrangement of twelve data visualizations arranged from least complex, to most complex, based on the complexity scores obtained from the study. THe visualizations are: piechart, normalized stacked barchart with a good legend and description, a 3D piechart with many categories, a linechart with two data-series in the smae plot area, a grouped barchart where every bar is their own colour, two barcharts and two linecharts superimposed on top of each other in the same plot area, a 3D grouped barchart, a treemap, a 3D bubbleplot, a node-link diagram which encodes hierachical data in each node, a densely connected node-link diagram, and finally a parallel coordinate plot.},width=0.95\linewidth, alt={A collage of visualizations arrange left to right, from simple to complex, based on the crowdsourced perceived complexity scores.}]{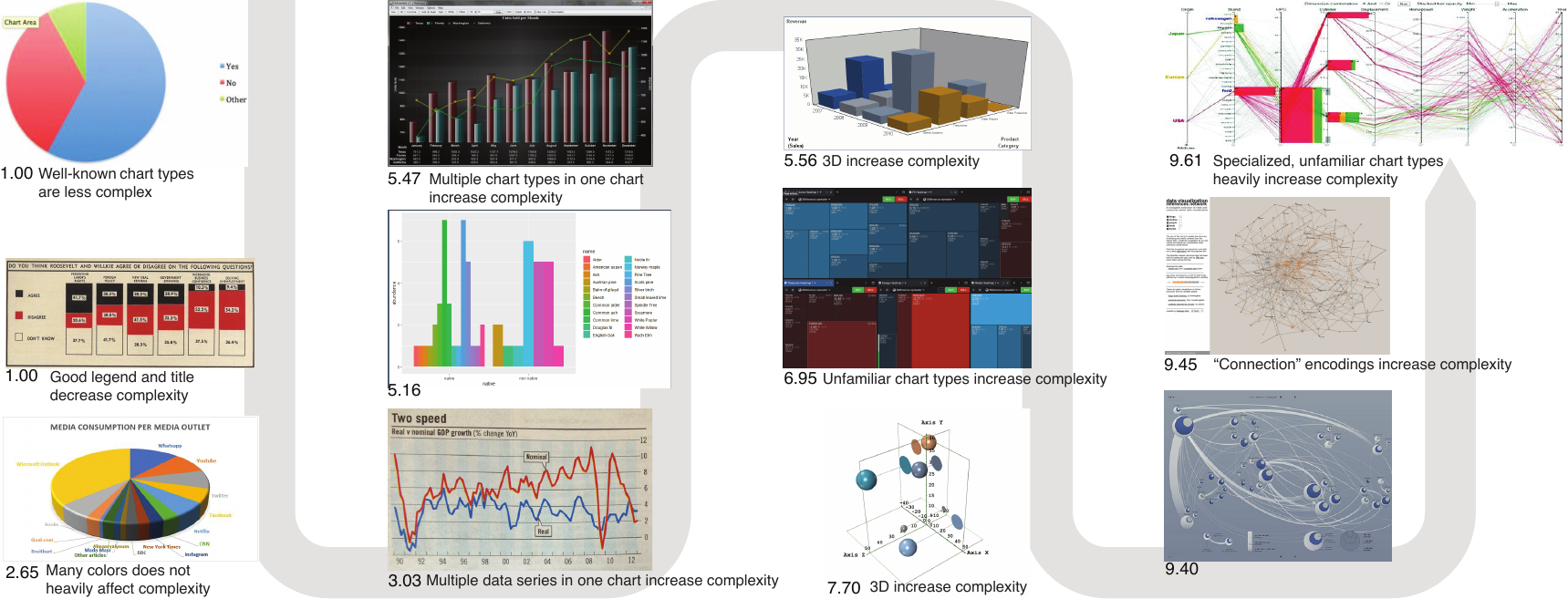}
  \caption{\textbf{Example visualizations from our dataset.}
  The charts are arranged from least complex (left), to most complex (right) based on the perceived complexity scores collected through the crowdsourcing study.
  Simple visualizations include bar- and piecharts, as well as visualizations with sufficient legends and descriptions. 
  Complex visualizations include node-link diagrams, parallel coordinate plots, and treemap diagrams, as well as composition layouts, such as small multiples, and multiple coordinated visualizations. 
  }
  \label{fig:teaser}
}

\graphicspath{{figs/}{figures/}{pictures/}{images/}{./}} 

\usepackage{booktabs}                  
\usepackage{mathptmx}                  
\usepackage{fontawesome5}
\usepackage[table,xcdraw]{xcolor}
\usepackage{comment}
\usepackage{subcaption}

\usepackage{wrapfig}

\usepackage{soul} 
\definecolor{new-green}{rgb}{0.104,0.667,0.229}
\newcommand{\add}[1]{{#1}} 
\newcommand{\rev}[1]{{#1}} 
\newcommand{\del}[1]{} 

\begin{document}

\firstsection{Introduction}
\maketitle

\input{sections/01-introduction}
\input{sections/02-related_work}
\input{sections/03-vis-complexity}
\input{sections/04-method}
\input{sections/06-discussion}
\input{sections/07-conclusion}
\input{sections/acknowledgements}

\bibliographystyle{abbrv-doi-hyperref} 
\bibliography{beholder}

\end{document}

%% file: sections/abstract.tex
Constructing expressive and legible visualizations is a key activity for visualization designers.
While numerous design guidelines exist, research on how specific graphical features affect perceived visual complexity remains limited.
In this paper, we report on a crowdsourced study to collect human ratings of perceived complexity for diverse visualizations.
Using these ratings as ground truth, we then evaluated three methods to estimate this perceived complexity: image analysis metrics, multilinear regression using manually coded visualization features, and automated feature extraction using a large language model (LLM).
Image complexity metrics showed no correlation with human-perceived visualization complexity.
Manual feature coding produced a reasonable predictive model but required substantial effort.
In contrast, a zero-shot LLM (GPT-4o mini) demonstrated strong capabilities in both rating complexity and extracting relevant features. 
Our findings \rev{suggest} that visualization complexity is truly in the eye of the beholder, yet can be effectively approximated using zero-shot LLM prompting, offering a scalable approach for evaluating the complexity of visualizations. \add{The dataset and code for the study and data analysis can be found at \url{https://osf.io/w85a4/}.}

%% file: sections/01-introduction.tex

Some visualizations are more difficult to interpret than others.
A node-link diagram of a social network, for instance, demands more of the reader to interpret than a bar chart of quarterly sales.
This difference suggests an inherent \textit{complexity} in visualizations---both an objective measure of the cognitive effort required to extract insights from a visual representation of data, as well as the subjective perceived complexity of the visualization.
But it also raises a fundamental question: can we quantify and correlate these complexities?
More specifically, can we derive an objective complexity metric based solely on graphical features: the lines, points, and bars; their color, thickness, and texture; the labels, titles, and legends? 
And what is the connection between these features, and the perceived complexity? 
One possibility is to use standard image complexity measures from computer vision---such as Shannon's entropy, image complexity~\cite{mahon_minimum_2024}, and other perceptual metrics---to derive the complexity of a visualization. 
Or is perhaps visualization complexity intrinsically tied to the human observer? 
Obviously, the human visual system has evolved to favor certain visual characteristics over others: seeing trends, anomalies, patterns, clusters, and gestalts.
And how does this novel complexity notion interplay with the more established one of \textit{visualization literacy}~\cite{DBLP:journals/tvcg/BoyRBF14, chevalier_observations_2018, borner_investigating_2016, borner_data_2019, organ_data_2024}, which measures the capability of the human viewer to understand a visualization?
In other words, how much of a visualization's complexity lies in the eye of the beholder?

In this paper, we explore this novel concept of visualization complexity through three complementary experiments. 
We begin 
with a crowdsourced study involving \rev{318} participants to collect the perceived visualization complexity rating of 640 visualizations \add{gathered from the internet} on a scale from 1 to 10.
Using this dataset as a ground truth of perceived complexity, we structured our investigation as follows:

\begin{enumerate}
    \item[E1] \textbf{Experiment 1} investigates if image complexity metrics drawn from computer vision---such as entropy---correlate with the perceived visualization complexity from the crowdsourced data. 

    \item[E2] \textbf{Experiment 2} performs a multilinear regression analysis with manually extracted features to investigate how different features of the visualizations affect the perceived complexity.

    \item[E3] \textbf{Experiment 3} examines the use of the vision-enabled LLM GPT4o-mini for both the automatic extraction of features in visualizations, and for directly rating visualization complexity, and compares it to manual coding and the crowdsourced data. 
\end{enumerate}

Our results indicate that standard computer vision techniques are unsuited as metrics for the perceived complexity of visualizations.
On the other hand, we find that a multilinear regression model can be used to estimate the perceived complexity ratings, with relatively large deviations.
Of course, this requires time-consuming manual coding of visualization features, but does provide some insight into how aspects of a visualization contribute to its perceived complexity.
Finally, we find that vision-enabled LLMs are capable of extracting most features from a visualization, even with a zero-shot prompt, meaning that the cumbersome task of manually extracting features may not be required. 
Likewise, LLMs are able to successfully estimate the human-rated complexity of the visualizations. 

\add{These results are, however, not conclusive and more work is needed to understand what affects the complexity of visualizations. 
Our dataset had a wide but not exhaustive variety of visualizations of different quality, which means that visualizations were not evenly represented, and some common ones, like stacked area charts~\cite{lee_vlat_2017}, were not present}.

Our contributions in this paper are as follows (code for the study and experiments as well as supplemental material can be found at \url{https://osf.io/w85a4/}): 
(1) the definition of visualization complexity as a new metric that can be used to assess the difficulty of understanding a specific chart; 
(2) results from a crowdsourced study assessing perceived visualization complexity for some 640 charts; 
(3) an investigation of the correlation between objective image complexity metrics, visualization features, LLM interpretation, and perceived visualization complexity; and 
(4) a robust discussion of concrete approaches for measuring visualization complexity as well as the pros and cons of pursuing further work in visualization complexity. 

%% file: sections/02-related_work.tex
\section{Background}

This paper draws on prior research in graph comprehension, visualization literacy, and crowdsourcing studies within visualization research. 
Here we review the relevant literature on these topics.

\subsection{Graphical Perception}
\label{RW-graph-perception}

Graphical perception has long been a topic of interest in the statistics community. 
Early work by Cleveland and McGill~\cite{cleveland_graphical_1984, cleveland_experiment_1986, cleveland_graphical_1987} systematically examined the effectiveness of different visual encoding channels. 
They rank ten different encodings based on their expressiveness for reading values of a visualization. 
Mackinlay built on this to rank encodings by effectiveness for different data types with the goal of automating chart construction~\cite{mackinlay_automating_1986}. 
Stewart and Best~\cite{stewart_examination_2010} later comprehensively examined Cleveland and McGill's graphical perception theory, and suggested that the original ten rankings should be no more than four. 

This prior literature focused on the effectiveness for reading values of a chart, but do not take into consideration how easy the chart is to interpret for the viewer. 
Taking their findings at face value, we would expect that at scatterplot would be among the most widely used visualizations, but this is not the case~\cite{lee_vlat_2017, boucher_mapping_2024, borkin_what_2013}. 
This suggests that other aspects are at play when authors choose which visualization to use, beyond encoding effectiveness for information extraction tasks. 

Edwards Tufte's data-ink ratio~\cite{tufte_visual_1983} was originally 
proposed as a measurement of how effectively people can draw conclusions from a visualization~\cite{inbar_minimalism_2007}.  
Tufte's idea was that if a chart shows only the data and no ``chart junk'' then it will be more easy to draw conclusion from the data~\cite{tufte_data-ink_1990}. 
However, optimizing the data-ink ratio alone does not guarantee that the visualization is simple and effective~\cite{inbar_minimalism_2007, mcgurgan_graph_2021}. 
Changing existing chart designs to minimize chart junk might also be detrimental, since people have learned the previous design, or schema, and will thus be unfamiliar with the new design. 
Tufte's principle of minimizing chart junk transforms the graphical primitives~\cite{bertin_semiology_1983} used in traditional designs.
For example, a box and whiskers plot typically composed of lines and areas is reduced in Tufte's approach to just two lines and a dot
This reduction in primitive elements reflects Tufte's underlying assumption that visualizations with fewer, more basic graphic primitives will be more effective---a hypothesis that warrants examination in the context of perceived complexity.

\subsection{Graph Comprehension and Construction}
\label{RW-graph-comprehension-and-construction}

Pinker introduced his Theory of Graph Comprehension \cite{pinker_theory_1990}, where he breaks down graphs into hierarchical schemas of structures, encodings, and the messages contained within these encodings.
His \textit{Graph Difficulty Principle} states that the more viewers must decode a graph from top down---rather than recognizing familiar conceptual messages encoded in the graph schema—the more time and memory consuming this task becomes.
This principle directly relates complexity to performance time in information extraction tasks.
According to Pinker, complexity can be reduced by using either simpler schemas or, importantly, schemas the viewer is already familiar with.
This suggests that perceived complexity is inseparable from a viewer's visualization literacy---a relationship we examine in our current study.

Carpenter and Shah~\cite{carpenter_model_1998} proposed a related model where graph comprehension emerges as a sequential process from pattern recognition to meaning construction.
Halford et al.~\cite{halford_how_2005} established that humans can process only four variables simultaneously without chance performance, providing an important cognitive constraint for visualization complexity.
Building on these foundations, Rodrigues et al.~\cite{rodrigues_spatial-perceptual_2007} offered a three-dimensional spatial-perceptual design space (shape, color, and spatialization) that frames visualization comprehension beyond pre-attentive processing models.

Work on measuring visualization effectiveness also informs our understanding of complexity.
Cabouat et al.~\cite{cabouat_previs_2024} introduced PREVis, a technique that measures perceived readability of visualizations using an 11-question assessment across four dimensions.
This structured approach shares conceptual ground with our attempt to measure complexity.
Haehn et al.~\cite{haehn_evaluating_2019} evaluated convolutional neural networks (CNNs) for predicting human graphical perception and found that, despite their capabilities, CNNs were not good models for human perception of visualizations---a caution we heed when comparing computational metrics with human judgments.

Graph comprehension can also be generative.
Haass et al.~\cite{haass_modeling_2016} modeled human comprehension to provide methods for constructing expressive visualizations. 
Tools such as Voyager~\cite{wongsuphasawat_voyager_2016} and its successor Voyager 2~\cite{wongsuphasawat_voyager_2017} recommend ``good'' visualizations based on perceptual ranking scores derived from classic graphical perception research~\cite{cleveland_graphical_1984, mackinlay_automating_1986, bertin_semiology_1983}.
\rev{Similarly, Draco~\cite{DBLP:journals/tvcg/MoritzWNLSHH19}, Draco 2~\cite{yang_draco2_2023}, and DracoGPT~\cite{wang_dracogpt_2025} employ constraint solving with theoretical design knowledge to recommend visualizations.}
These systems operationalize aspects of visualization quality, but typically focus on perceptual effectiveness rather than complexity.
Our work complements these approaches by specifically examining how to measure the complexity dimension of visualizations, which could potentially enhance such recommendation systems.

\subsection{Visualization Literacy}

Visualization literacy is a newer term than graph comprehension, though its definition varies across the literature.
Ge et al.~\cite{ge_toward_2024} note the lack of a unified definition, but most characterizations include an individual's ability to read and interpret data visualizations~\cite{chevalier_observations_2018, lee_vlat_2017}.
Some definitions extend this to include interpreting patterns, extracting meaning, and using visualizations to answer data-related questions~\cite{borner_investigating_2016, DBLP:journals/tvcg/BoyRBF14, borner_data_2019}.
Firat et al.~\cite{firat_interactive_2022} provide a comprehensive survey of this domain, highlighting the need to study cognitive impacts of visualizations---a direction that aligns with our complexity research.

Efforts to measure visualization literacy have produced several assessment tools.
Lee et al.~\cite{lee_vlat_2017} developed VLAT (Visualization Literacy Assessment Test), defining visualization literacy specifically as ``\textit{the ability and skill to read and interpret visually represented data in and to extract information from data visualizations}''~\cite[p. 552]{lee_vlat_2017}.
This narrow definition enabled quantifiable metrics for assessment.
Building on this work, Pandey \& Ottley~\cite{DBLP:journals/cgf/PandeyO23} introduced Mini-VLAT, a shortened version of the test. 
Taking a different approach, Ge et al.~\cite{DBLP:conf/chi/GeC023} extended visualization literacy to include critical thinking, developing CALVI to assess critical visualization comprehension skills.

Research on improving visualization literacy spans various demographics and visualization types.
Programs target children~\cite{chevalier_observations_2018, bae_cultivating_2023, DBLP:conf/eurographics/FiratDL20}, adults across all ages~\cite{organ_data_2024}, and STEM students~\cite{costa_graphical_2020}.
Specific visualization types have received focused attention, including scatterplots~\cite{wang_what_2022}, treemaps~\cite{DBLP:conf/eurographics/FiratDL20}, and parallel coordinate plots~\cite{DBLP:conf/chi/KwonL16, stoiber_visualization_2019}.
Recent research has also explored using large language models to assist with chart interpretation~\cite{choe_enhancing_2024}.

Broader visualization literacy research examines visualization onboarding strategies~\cite{stoiber_visualization_2019, stoiber_design_2021, stoiber_abstract_2022, stoiber_design_2023, lee_how_2016, ruchikachorn_learning_2015}, and existing challenges in the field~\cite{bach_challenges_2024, nobre_reading_2024}.
These efforts highlight the importance of considering how people learn to interpret unfamiliar visualizations.

While our paper does not directly measure graph comprehension, a reader's existing visualization literacy is important when assessing perceived complexity.
A visualization that leverages schemas and encodings familiar to the reader will likely be perceived as less complex than one introducing entirely novel representations.
This relationship between literacy and perceived complexity informs our experimental design and analysis.

\subsection{Crowdsourcing Studies in Visualization Research}

Crowdsourcing has become an established method for visualization research since Heer and Bostock's seminal work~\cite{DBLP:conf/chi/HeerB10}, which demonstrated that crowdsourced studies could replicate laboratory results from Cleveland and McGill~\cite{cleveland_graphical_1984} with acceptable variation at significantly lower costs.
This approach has proven particularly valuable for graphical perception studies.

Building on this foundation, researchers have applied crowdsourcing to various visualization evaluation contexts.
Sasaki and Yamada~\cite{sasaki_crowdsourcing_2019} used crowdsourcing to investigate contrast in visualization, while Turton et al.~\cite{turton_crowdsourced_2017} employed it for colormap assessment.
Borgo et al.\ conducted comprehensive reviews of crowdsourcing in visualization research~\cite{borgo_crowdsourcing_2017} and specifically for visualization evaluation~\cite{borgo_information_2018}, proposing standardized reporting formats to improve reproducibility.

For our study, we leverage these established crowdsourcing approaches using the ReVISit tool~\cite{ding_revisit_2023} to conduct our visualization complexity assessment. 
We adhere to the reporting framework from Borgo et al.~\cite{borgo_information_2018};  details can be found in our supplemental materials.

%% file: sections/03-vis-complexity.tex
\section{Visualization Complexity}

We here discuss our notion of \textit{visualization complexity} as a measure of the difficulty of understanding a visualization.
We start by unpacking the concept of complexity.
We then analyze how this concept has been applied to the visual domain, before we move to the realm of visualization. 
Finally, we derive a working definition of visualization complexity and how we might measure it.

\subsection{Complexity}

The concept of \textit{complexity} has been extensively studied by researchers. 
Heylighen~\cite{heylighen_growth_1999} describes complexity as a combination of variety and dependency. 
Variety refers to the fact that a complex system consists of multiple distinct parts, that behave differently.
Dependency refers to the fact that these parts are not independent, rather, knowledge of one part can tell something about another part. 
Symmetry is used as an example of a quality which increases order, and thus decreases complexity. In turn, complexity is characterized by the lack of symmetry, where no part of the system can provide enough information to predict any other part in the system. 
Heylighen points out that neither complete order nor complete chaos is characterized by complexity. 
Complete order is deterministic, while complete chaos is statistically homogeneous.  
Based on these points, we can derive the following definition:
\begin{quote}
    \textbf{Complexity:} A property that emerges from the combination of variety (multiple distinct parts with different behaviors) and dependency (interconnectedness between parts where knowledge of one part provides information about others).
\end{quote}
Complexity can also be seen through a subjective lens based on the perception of a viewer. 
In addition, the complexity of a system is related to the scale at which it is viewed. 
When viewed at an appropriate scale, what was complex at a smaller scale will seem simple \cite{heylighen_growth_1999}.

\subsection{Visual Complexity}

In the visual domain, prior work has examined which features constitutes \textit{visual complexity}.
Berlyne~\cite{berlyne1958influence, berlyne1970novelty} was among the first to systematically define visual complexity in terms of objective stimulus properties.
He emphasized variables such as the number of elements, irregularity, and heterogeneity as key contributors to complexity, proposing that these factors influence aesthetic response and cognitive arousal.
This focus on objective properties aligns with Heylighen's later conceptualization of complexity as arising from the interplay of variety and dependency between elements.
Berlyne’s framework shifted attention away from purely subjective judgments and toward quantifiable characteristics of the visual stimulus, establishing a basis for later empirical and computational approaches to measuring visual complexity.

Oliva et al.~\cite{oliva_identifying_2004} investigated the perceived visual complexity of scenes, where participants ranked the complexity of images of rooms from low to high.
They found that visual complexity is affected by the quantity of objects, clutter, openness, symmetry, organization, and the variety of colors. 
Reducing these features, individually or together, reduced the perceived complexity of the scene. 

Oliva et al.'s dimensions of symmetry and organization can be directly mapped to Heylighen's notion of symmetry: 
Both can be used to predict features of the scene, thus reducing complexity. 
The dimensions of quantity of objects, clutter, and variety of colors can be seen as a measure of the amount of entropy, or uncertainty, in the scene, again linked to the predictability of these objects' position and color. 
In a scene with many differently colored objects, in a disorganized, cluttered fashion, it is much harder to predict the structure of these objects, than if the scene contain few, neatly organized objects in consistent colors. 
Based on these points, we derive the following definition: 
\begin{quote}
    \textbf{Visual complexity:} The objective measure of intricacy within any visual stimulus, quantifiable through its fundamental compositional characteristics.
    This includes the number of discrete visual elements, their variability (in size, shape, color, orientation), the density of these elements, their spatial distribution and arrangement, the diversity of patterns, and the level of symmetry or irregularity present.
\end{quote}
As with the complexity, visual complexity can also be a subjective measure.
Much of vision science research is concerned with determining a relationship between such objective and subjective measures.

\subsection{Visualization Complexity}

Building on prior work in complexity, graph comprehension, and visualization literacy, we here conceptualize \textit{visualization complexity}. 
Going by the above argument, we would assume that a visualization is the least complex when it contains few and predictably placed elements, in a simple color scheme. 
However, good visualizations are also constrained by the fact that they must be information-rich, expressive, and truthfully represent the underlying data.
Even basic visualizations, such as a barchart, contains many visual elements~\cite{Wilkinson2005}. 

Applying Heylighen's notion of representational levels, visualization elements can be analyzed at different scales.
Just as an axis can be broken down into a line, with tick-marks, and numbers, each of which can be further broken down into their geometric shape, we can also abstract from these details, and consider the scales as a single entity. 
Similarly, the dots of a scatterplot can be seen as a number of individual dots, with seemingly little connection between the other dots, or they can be viewed as a group of dots that represent the distribution and density of some data values, as delimited by the axes.  

Some treatments are grounded in subjective perception.
As discussed previously, Pinker~\cite{pinker_theory_1990} considers visualizations with simpler schemas, especially those already familiar to viewers, to be more comprehensible.
In other words, in his treatment, the visualization literacy of the viewer is central to how the viewer perceives the complexity of a visualization, regardless of any objective metric. 

Few studies have attempted to measure subjective visualization complexity.
Zhu~\cite{zhu_measuring_2007} established principles for effective visualizations based on accuracy, utility, and efficiency, emphasizing reduced cognitive load.
Their analytical approach~\cite{zhu_complexity_2007} decomposes visualizations into hierarchical components and visual patterns, assigning complexity scores through heuristic rules.
This produces both a complexity tree structure and an overall score.
However, this approach requires expertise in visualization and cognitive psychology, is time-consuming, and produces abstract numerical scores difficult to interpret meaningfully.

In this paper, we restrict our definition to objective measurement of the complexity of non-interactive visualizations. 
Based on the above, we propose the following definition:  
\begin{quote}
    \textbf{Visualization complexity:} The structural and informational intricacy of a data representation, determined by its compositional attributes and informational architecture.
    This encompasses the number of data variables encoded, the types of visual encodings employed, the number of distinct marks or glyphs, the layers of information presented, the number of scales used, the diversity of visual channels utilized, and the relational structure between elements.
\end{quote}
One of our goals in this paper is to determine the relationship between objective and subjective measures of visualization complexity.

\subsection{Features of Visualization Complexity} 
\label{sec:vizCompl-features}

Several objective features of a visualization might influence its complexity. 
We review a synthesis of existing work below:

\begin{itemize}
    \item[\faDatabase] \textbf{Scale:} For most visualizations, the larger the dataset to visualize, the higher the resulting complexity.
    This is particularly true for unit visualizations~\cite{DBLP:journals/tvcg/ParkDFE18}, where there is a one-to-one mapping between data points and visual marks.

    \item[\faPalette] \textbf{Colors:} The number and choice of colors clearly may influence how busy a visualization seems; Zhu argues for this relation between color and data domain~\cite{zhu_complexity_2007}. 
    A busy background might also add to the perceived complexity of a visualization, as opposed to one that has a clear, single-colored background \cite{oliva_identifying_2004}. 

    \item[\faCube] \textbf{3D:} Three-dimensional visualizations add an extra spatial dimension to interpret. 
    This results in the viewer having to cognitively process the spatial structure of a 3D visualization, adding an extra dimension to the visualization schema~\cite{pinker_theory_1990}, in which elements can be organized~\cite{oliva_identifying_2004}.

    \item[\faSign] \textbf{Legends:} Even common visualizations can seem overwhelming without a sufficient description, labels, and legends. 
    A lack of a \textit{good} description, not just a description, or of a \textit{good} legend, would therefore likely have some effect on the complexity~\cite{cleveland_experiment_1986, munzner2014visualization}.

    \item[\faComment] \textbf{Title and annotations:} Textual annotations can help the viewer make sense of what she is seeing. 
    However, annotations can also confuse the viewer if they are inappropriate for the task the viewer tries to achieve, or if they are not properly explained~\cite{DBLP:conf/chi/KongLK18}. 

    \item[\faClone] \textbf{Visual overlap:} Several chart types have variants that stack or overlap or group elements. 
    This can facilitate comparison, and allows for more data series in the same chart, but also increases the amount of things going on in a single chart, complicating the schema of the visualization~\cite{pinker_theory_1990}, due to the increase in elements and possible clutter~\cite{oliva_identifying_2004}. 

    \item[\faPlusCircle] \textbf{Composition:} Compositions~\cite{DBLP:conf/apvis/JavedE12} of multiple different chart types---e.g. a bar chart with a  superimposed line chart---likely increases complexity; cf.\ Pinker~\cite{pinker_theory_1990, oliva_identifying_2004}. 
    The same is likely true with multiple data series, small multiples, or coordinated views. 
\end{itemize}

\subsection{Measuring Visualization Complexity}

Our research focuses specifically on the connection between the perceived complexity of visualizations and the objective complexity of the visualizations' features at different levels of abstraction.
These range from low-level graphical primitives and encoding channels to high-level features such as color variety, multiple scales, 3D effects, chart type combinations, and spatial characteristics like space-filling or overlapping layouts.
In the following section, we empirically evaluate several practical approaches to measuring visualization complexity.

%% file: sections/04-method.tex
\section{Study and Experiments}

Here we detail our crowdsourcing study and the experiments we conducted to correlate visualization complexity with perceived complexity.

\subsection{Crowdsourcing Perceived Visualization Complexity}
\label{sec:crowdsourcing}

We initially collected a dataset of perceived complexity scores of various visualizations using a crowdsourced study. 
These scores were treated as the ground truth of the visualization's perceived complexities. 
Below we detail the study in more details.
A condensed experiment reporting form~\cite{borgo_information_2018} is provided as supplemental material. 

\subsubsection{Stimulus}

We used the dataset of visualizations from Shin et al.~\cite{DBLP:journals/tvcg/ShinCHE23}. 
The original dataset includes 12,487 images of visualizations collected from the internet. 
We noticed that the dataset contained some duplicates, so we removed duplicates by hashing the images, and comparing these for duplicates. 
We randomly sampled 640 visualizations from the dataset using the ReVISit tool~\cite{ding_revisit_2023} used to conduct the study. 
See figure \ref{fig:teaser} for some examples of the stimulus used. 
All visualizations were {\scshape jpg} image files. 

Since the dataset is collected from the internet, there is a large difference in quality of the visualizations, and the quality of the image files themselves. 
The images contain visualizations ranging from example visualizations of barcharts, to infographics with rich storytelling features, to screenshots of interactive tools that have performed some selection or filtration of the data. 
In addition, the visualizations have various amounts of styling applied, and some visualizations consist of multiple coordinate visualizations (MCV) in a dashboard layout, or visualizations superimposed on top of each other, e.g., a linechart superimposed on top of a barchart. 
Due to these features, some visualizations also included multiple y-axes, are 3-dimensional in nature, or appear to visualize multiple datasets in consort. 
Several visualization have axes, legends, titles, and descriptions cut off in the image, or be of a quality too low to read the description. 
\add{This variety of visualizations, including their quality, helps the dataset reflect the kind of visualizations one might encounter in the wild, as opposed to only perfectly crafted visualizations made using the state of the art authoring tools.}

The 640 visualizations sampled for the study contain 51 different visualization types,
\add{with which we mean distint, named visualizations, like barchart, stacked barchart, scatterplot, bubbleplot, etc.}
\rev{The most represented chart type in the dataset is barcharts}, with 132 visualizations, followed by linecharts (87), piecharts (52), candlestick charts (46), and areacharts (38). 
11 visualization types were only represented once in the sample dataset. 
A full breakdown of the sampled visualizations can be found in the supplemental material.


\rev{
We used reVISit's builtin \textsc{latinsquare} with subsampling, to sample 50 visualization for each participant. 
}

\subsubsection{Task}

The task consisted of asking the participant to rate their initial impression of the complexity of a visualization shown in their web browser on a Likert scale from 1 to 10 (1 being simplest, and 10 being most complex).
All options were labeled. 
There was no time limit on this rating task.
\add{Participants were also asked to provide at least one reason for why they gave the visualization this rating using a multi-select tag picker.}
Once the participant had completed the task, they would click ``next'' to proceed.
A full session constituted rating a total of 50 visualizations per participant. 
\add{There were no attention checks included in the task.}

\subsubsection{Participants}

We recruited a total of 625 participants through the Prolific crowdsourcing platform.
\rev{309 participants} left the study without completing it, 27 participants timed out. 
We accepted the remaining 289 participants.
Participants who completed more than 50\% of the session were included in the data, and were compensated for their time. \add{This resulted in 318 participants in the study.}
Participants were required to complete the task on a computer or tablet, excluding participation on smartphones, due to the small size of such devices. 

Participants were compensated £8.92/hr, as per Prolific's recommended £9/hr.
The task took on average 32 minutes, resulting in £4.83 in compensation to the participants. 
The slightly lower pay is due to some participants taking more time than estimated for the task, but is still well above Prolific's minimum payment of 6£/hr. 

Participants were prescreened to be fluent in English, and the quota was set to be 50\% female. 
Their ages ranged from 19 to 72 years of age (M=29.8, SD=9.3). 
\rev{30\% of the participants had a bachelor's degree, and 47\% had a lower education level than a bachelor's degree}. 
50\% of the participants reported that they encountered data visualizations a few times a month or less in \emph{private or personal} contexts. 
56\% of the participants reported that they encountered data visualizations a few times a month or less in \emph{professional} contexts.
For both private and professional context, 14\% of the participants reported encountering visualizations a few times a week, and 13\% of them encountered visualizations several times a week. 
The exposure to visualizations in both private and professional contexts were otherwise very similar. 

\add{Participants were distributed around the globe, with the largest country of origin being South Africa (120), followed by Portugal (35), and Poland (34). Participants were predominantly from Europe and South Africa.  Table~\ref{tab:participant-origin} gives an overview of the origin of the participants per region.}

\begin{wraptable}{r}{0.46\columnwidth} 
\caption{\textbf{\add{Origin of participants}}.}
\label{tab:participant-origin}
\begin{tabular}{@{}lrr@{}}
\toprule
Origin        & Count & \% \\ \midrule
Africa        & 122   & 38.4       \\
Asia          & 5     & 1.6        \\
Europe        & 168   & 52.8       \\
Middle East   & 1     & 0.3        \\
North America & 14    & 4.4        \\
Pacific       & 3     & 0.9        \\
South America & 5     & 1.6        \\ \midrule
Total         & 318   & 100        \\ \bottomrule
\end{tabular}%
\end{wraptable} 

\subsubsection{Procedure}

The study was posted on Prolific with a description of the study's purpose, the involved tasks, and the provided compensation.
Participants were first asked for their informed consent, and afterwards instructed to fill in a demographics form. 
The participants where then introduced to the task in more detail, including that they should answer their immediate `gut feeling' and that the visualizations might be missing details like legends or descriptions. 
They were then introduced to the study user interface, and 
finally the participants were presented with six training examples, to further familiarize them with the task. 
The participants could see the task instructions and example visualizations at any time by pressing the help button in the ReVISit user interface. 
Afterwards, the participants began the task. 
Upon completion, the participants were provided with a link to return to Prolific, and mark their participation in the study as completed. 
The study took on average 32 minutes and 30 seconds to complete. 

\subsubsection{Data Collection Platform} 

We built our crowdsourced data collection as a web application using the ReVISit tool~\cite{ding_revisit_2023}. 
We chose Prolific as the crowdsourcing platform because of its excellent support for HCI research and \rev{high-quality crowdworkers~\cite{crowdsourcing-quality}.}
Data was stored using Google Firebase. 

\subsubsection{Measurements and Metrics}

The measurements for the study was the self-reported perceived complexity rating. 
This results in each visualization being associated with a list of complexity ratings. 
There was no performance metrics being measured, such as completion time, but the information is automatically saved by the ReVISit tool for each task. 
The subsequent experiments performed all center around the collected perceived complexity score.
\add{
Due to an error in our ReVISit configuration, each visualization was not sampled an even number of times, and the tags the participants were asked to input were each unique, resulting in 3754 different tags. We therefore decided not to analyze the tags for this paper. 
}
We report on the experiments in the following sections. 

\subsubsection{Hypotheses}

We formulate the following hypotheses for the crowdsourced data: 

\begin{enumerate}
    \item[\textbf{H1}] \textit{\rev{Commonly used charts} will all have among the lowest complexity scores.}
    \rev{These chart types are familiar to most people, and this familiarity will reduce the perceived complexity. These chart types are among the most used in curricula~\cite{lee_vlat_2017,boucher_mapping_2024}, news outlets~\cite{lee_vlat_2017, borkin_what_2013}, and in visualization authoring tools~\cite{lee_vlat_2017}. We refer to the 12 visualizations used in VLAT~\cite{lee_vlat_2017}, which are based on these criteria.}
    
    \item[\textbf{H2}] \textit{Annotations reduces complexity.} 
    Labels and textual descriptions presumably help explain a visualization.
    
    \item[\textbf{H3}] \textit{Composition increase complexity.}
    Multiple scales, composing multiple chart types, or using multiple views will increase the perceived complexity of the visualization. 
    
    \item[\textbf{H4}] \textit{Legends reduce complexity.} 
    A good legend will help explain a visualization, thereby reducing its complexity.
\end{enumerate}


\subsubsection{\add{Results}}

The participants from our crowdsourcing study were relatively inexperienced in the use of data visualizations, as the self-reporting indicates.  
The majority of the participants (50\%) only encountered visualizations a few times per month, so it is not unreasonable for them to be unfamiliar with many of the more specialized chart types, such as node-link diagrams, and alluvial diagrams. 
Few participants also experienced visualizations often, which means that the participants can be considered as lay-people in the field of data visualizations. 

The chart type with the lowest average perceived complexity score was pictograms; however, there was only one example of this visualization in the dataset. 
The second least complex visualization was pie charts. 
This is not surprising, as piecharts are a common visualization type that we expect most of the participants to have encountered before. 
\rev{Of the common chart types employed in VLAT~\cite{lee_vlat_2017}, piechart (2.16, SD=0.90), barchart (3.36, SD=1.70), histogram (3.40, SD=1.58), choropleth map (4.18, SD=1.45), linechart (4.38, SD=2.12), and areachart (4.74, SD=2.10) are among the least complex half of the visualization types in the dataset. The first three are also among the 10 least complex visualization types. Bubblechart (5.80, SD=2.13), scatterplot (6.43, SD=2.21), and treemap (6.85, SD=2.38) were rated significantly higher in terms of perceived complexity. The remaining charts from VLAT were not present in the dataset.}
This is in line with our expectations in hypothesis \textit{H1}. 

The most complex visualization type was parallel coordinates. 
This makes sense from the work of Pinker~\cite{pinker_theory_1990}, in that a parallel coordinate set does not resemble any of the \rev{common chart types} most viewers are somewhat familiar with in terms of schema and encoding\rev{~\cite{lee_vlat_2017, borkin_what_2013, boucher_mapping_2024}.} 
Therefore, there is little to no point of origin to begin to understand how the visualization works, and for a study like ours with no-one to ask for guidance, the visualization remains cryptic. 
Other visualization types with a high average complexity score were node-link diagrams, arc charts, alluvial diagrams, chord diagrams, connected scatterplots and candlestick charts. 
For most of these visualization types, the existing knowledge of visualization schemas and encodings from common chart types are likewise not easily applied. 
An exception is the candlestick chart, which could be perceived as a linechart with extra information encoded, but these visualizations often had domain specific annotations drawn on top of them, likely causing the increase in perceived complexity.
We return to annotations in the discussion. 

\add{
It is important to note the large variety in the scores for a given visualization type. E.g. different barcharts were rated everything from 1 to 10, meaning that chart type alone cannot explain the perceived complexity ratings of the visualizations. 
}
In summary, the results from the study suggests that the perceived complexities of visualizations depend on the familiarity with the schema, inline with Pinker's principle~\cite{pinker_theory_1990}. 

\subsection{Experiment 1: Image Analysis}

Our first experiment involved measuring the complexity of a visualization using objective metrics rather than  manual expert heuristics, such as is done by Zhu~\cite{zhu_complexity_2007}.
More specifically, this initial experiment was guided by the notion that the perceived complexity might be linked to the (dis)order of the visualization. 
In other words, the more structured a visualization is, the less complex it is perceived by the viewer. 
Furthermore, this structured-ness might be captured using existing visual complexity metrics that analyze images on a pixel-level, rather than the semantic level of the visualization components. 
Based on this notion, we added the following hypothesis: 

\begin{itemize}
\item[\textbf{H5}] \textit{There is a correlation between the image's pixel-based complexity, and the perceived visualization complexity.} 
\end{itemize}

\subsubsection{Method}

We investigated four specific visual complexity metrics:

\begin{enumerate}
    \item \textbf{Shannon entropy:} The average amount of unpredictability in a system by quantifying how many bits of information are needed to describe its possible states and their probabilities.
    This entropy metric was calculated using the \texttt{scipy} Python library.

    \item \textbf{Meaningful image complexity:} This metric, adapted from Mahon and Lukasiewicz~\cite{mahon_minimum_2024}, evaluates the structural complexity of a visualization by analyzing its core visual elements and their relationships rather than relying on raw pixel-level calculations.
    We used the code provided by Mahon and Lukasiewicz. 

    \item \textbf{Feature congestion:} Rosenholtz et al.~\cite{rosenholtz_feature_2005} suggest using a feature congestion metric of clutter for visualizations. They argue that high levels of clutter can be a detrimental quality for some tasks. 
    We use the provided Matlab code. 

    \item \textbf{Subband entropy:} Rosenholtz et al.~\cite{rosenholtz_measuring_2007} also suggest that a steerable pyramids based subband entropy measure can capture the clutter of an image, interface or visualization. 
    We use the provided Matlab code. 

\end{enumerate} 

We analyzed the results with Pearson's correlation analyses between the perceived complexity score and each visual complexity metric. 



%
%

\subsubsection{Results}
\label{experiment1-results}



The Pearson's correlation analyses showed 
(1) a negligible positive correlation between Shannon's entropy and the perceived visualization complexity (r = 0.019, p = 0.6), indicating no meaningful relation. This was expected. 
(2) a weak positive correlation between meaningful image complexity and perceived visualization complexity (r = 0.211, p < 0.01), contrary to our expectations.
(3) a negligible positive correlation between feature congestion~\cite{rosenholtz_feature_2005} and perceived visualization complexity (r = 0.007, p = 0.85). However, feature congestion~\cite{rosenholtz_feature_2005} was between 1 and 10 for all but one image, which was at 320. Ignoring this one extreme outlier, the correlation is weak (r = 0.211, p < 0.01).
(4) a weak positive correlation between subband entropy and perceived visualization complexity (r = 0.185, p < 0.01). 
With weak correlations, we must conclude that we cannot use either the meaningful image complexity, Shannon's entropy, feature congestion, or subband entropy to estimate a visualization's perceived complexity. 

\subsection{Experiment 2: Multilinear Regression Model}

Our next experiment involved deriving a multilinear regression model based on manually coded visualization features. 
The goal of this experiment is to understand how different features of a visualization affects the perceived complexity. 
While this at first glance might seem very similar to Zhu~\cite{zhu_complexity_2007}, we are not doing heuristic judgements on the different visual parts of the visualization, and using this to directly estimate the visualization complexity score. 
Rather, we code objective features in the visualization, such as chart type, number of colors, the presence of an appropriate legend and other features discussed previously. 
We do not pass judgment on the qualities of these features, e.g. that position is more expressive than area, or that certain chart types are simpler than others, but use them as dimensions in our multilinear regression analysis. 

\subsubsection{Method}
\label{sec:exp2-method}

We analyzed and coded each of the 640 visualizations in the study based on the below parameters. 
These parameters are an operationalized version of the complexity features discussed in section \ref{sec:vizCompl-features}. 
The coding was performed by one of the authors. 
While some of these features can be hard to code for some visualization, such as if the description and legend is sufficient, we relied on the author's expertise in estimating the most truthful answer (see supplemental material for the coding guide):

\begin{description}
    \item[Chart type:] The chart type used in the visualization. 
    \item[Stacked, overlapping, or grouped:] If the chart type is a stacked, overlapping, or grouped variant. 
    \item[Number of colors:] The estimated number of colors used in the data visualization.
    Colors in the background are not counted. 
    \item[Number of data series:] The number of data series the visualization shows; e.g.\ a line chart might show data from two or more data series, one per line. 
    \item[Data type:] An estimation of the data type; e.g.\ nominal, ordinal, discrete, time series, or continuous data. 
    \item[3D visualization:] Whether the visualization is a 3D visualization, including 3D effects.
    \item[Sufficient description:] Whether the visualization has a sufficient description and legend to be understood without further explanation of what is shown in the visualization. 
    \item[Single color background:] Whether the background is a single, solid color (as opposed to an image or other graphic). 
    \item[Multiple coordinated views:] Whether the visualization consists of multiple charts showing aspects of the same data. 
    \item[Small multiples:] Whether the visualization uses small multiples. 
    \item[Chart composition:] Whether there are multiple chart types in the same visualization; e.g.\ a line chart superimposed on a bar chart. 
    \item[Multiple scales on one axis:] Whether there are multiple scales used for a singe axis. 
    \item[Annotations:] Are any annotations present on the visualization, that provides some additional information beyond the base chart. 
\end{description}

Since the coding resulted in 50 different chart types being found in the dataset, we added additional coding to each visualization in the form of graphic primitives, and visual encoding channels used.  
The graphic primitives coding is based on Bertin's visual marks~\cite{bertin_semiology_1983}, and later additions to his work~\cite{borner_data_2019}, and consists of the following categories: 
(1) point,
(2) line,
(3) area,
(4) surface,
(5) volume,
(6) symbol.
If a visualization uses points as one of its graphical primitives, then this category is marked as 1 in the coding, otherwise it is marked as 0. 
A visualization can consist of multiple primitives. 

In addition, we also coded the visual encoding used by the visualizations. 
This was based on the work of Cleveland and McGill~\cite{cleveland_graphical_1984}, as well as Mackinlay~\cite{mackinlay_automating_1986}, who identify 13 visual encodings.
Similarly to the graphics primitives, a visualization can use more than one visual encoding channel.
If a channel is used for a visualization, we encode it as 1, otherwise as 0. We refer to Mackinlay~\cite{mackinlay_automating_1986} for a list of the visual encodings used.
The coding of graphical primitives and encodings was automated, based on the chart types of the visualization. \add{E.g. a barchart is an area primitive, and uses height for the encoding of the data.}
This coding scheme resulted in 32 features coded for each visualization. 

We used the final coding to perform a multilinear regression analysis with the perceived complexity score as the dependent variable and all manually coded features as independent variables.
We experimented with different sets of independent variables in the analysis; e.g. not taking into account the visual encoding channels, or not taking into account the number of colors used, etc. to find the best performing model. 
The goal was to understand if and how there is a correlation between the various coded features, and the perceived complexity score. 
With this experiment we expected the model to reflect hypotheses \textbf{H1--H4} from the crowdsourcing study. 
We investigate this in the discussion. 

\subsubsection{Results}


Of the different independent variables tried, the best performing model was with taking all features into account. 
\rev{The correlation value and weights for this model is shown in table \ref{tab:MRL-model}.} 
Note that a positive score means that the feature increases complexity, while a negative score means the feature reduces complexity. 
We return to this result in the discussion section, where we analyze and discuss it in detail. 
Based on the results from the multilinear regression analysis, we conclude that our multilinear regression model is reasonably able to capture the overall complexity of a visualization.
However, the model has significant variation in performance, and $R^2$ is quite low, at 0.49. 

\begin{table}
    \centering
    \caption{\rev{\textbf{Multilinear regression model.} 
    The last two rows in the right column show the bias and the coefficient of determination for the model.}}
    \label{tab:MRL-model}
    \begin{subtable}[t]{0.48\columnwidth}
        \resizebox{\textwidth}{!}{%
            \begin{tabular}[t]{@{}lr@{}}
                \toprule
                \textbf{Feature}                & \textbf{Weight} \\ \toprule
                connection$_{encoding}$         & 2.020  \\
                Small multiples                 & 1.516  \\
                volume$_{visual~primitive}$     & 1.509  \\
                volume$_{encoding}$             & 1.509  \\
                position$_{encoding}$           & 1.501  \\
                color saturation$_{encoding}$   & 1.402  \\
                symbol$_{visual~primitive}$     & 1.333  \\
                3D viz or 3D effect             & 1.212  \\
                Multiple scales                 & 1.144  \\
                Multiple charts                 & 0.975  \\
                Multiple chart types            & 0.805  \\
                containment$_{encoding}$        & 0.720  \\
                stacked or overlapping          & 0.698  \\
                line$_{visual~primitive}$       & 0.647  \\
                Annotations                     & 0.599  \\
                area$_{encoding}$               & 0.581  \\
                length$_{encoding}$             & 0.232  \\
                 \bottomrule
            \end{tabular}
        }
    \end{subtable}
    \hfill 
    \begin{subtable}[t]{0.465\columnwidth}
        \resizebox{\textwidth}{!}{%
            \begin{tabular}[t]{@{}lr@{}}
                \toprule
                \textbf{Feature}                & \textbf{Weight} \\ \toprule
                color hue$_{encoding}$          & 0.176  \\
                Plain background                & 0.175  \\
                overlapping layout              & 0.127  \\
                number of colors                & 0.090  \\
                Number of data series           & 0.042  \\
                density$_{encoding}$            & 0.000  \\
                texture$_{encoding}$            & 0.000  \\
                point$_{visual~primitive}$      & -0.191 \\
                area$_{visual~primitive}$       & -0.459 \\
                shape$_{encoding}$              & -0.888 \\
                Legend/description              & -0.933 \\
                slope$_{encoding}$              & -0.997 \\
                spacefilling layout             & -1.031 \\
                angle$_{encoding}$              & -1.241 \\
                surface$_{visual~primitive}$    & -1.347 \\
                \textbf{Model Bias}             & 4.364  \\
                \textbf{$R^2$}                  & 0.490  \\
                \bottomrule
            \end{tabular}
        }
    \end{subtable}
\end{table}


\subsection{Experiment 3: LLM-based Feature Extraction and Complexity Rating}

While our multilinear regression model yields acceptable performance, it requires manual---and potentially time-consuming---coding of visualizations.
In our third experiment, we were interested to see if we could use a vision-enabled large language model (LLM)---such GPT4o-mini---to extract features of a visualization as well as rate its complexity, in an automated fashion. 
Therefore, we formulated the following hypotheses on the performance of GPT4o-mini for extracting features of, and rating the complexity of, a visualization:

\begin{itemize}
\item[\textbf{H6}] \textit{GPT4o-mini is able to extract features from a visualization at a similar performance as an expert human.}
\item[\textbf{H7}] \textit{GPT4o-mini is able to estimate the same complexity of a visualization that a novice viewer would do.}  
\end{itemize}

\subsubsection{Method}

We designed two prompts; (1) extracting the above features of a visualizations (see Section~\ref{sec:exp2-method}; except for graphic primitives and visual encoding channels), and (2) rating visualization complexity  using the same score as in the crowdsourcing study (see Section~\ref{sec:crowdsourcing}). 

For extracting of features, we described each of the features, and asked the model to format the output in a predefined way. 
For rating visualization complexity, we asked for a number, as well as an argument for the rating it provided. 
Both prompts were zero-shot prompts, providing no examples of the complexity scores for the visualizations. 
Each prompt included a context message to the system, which was sent with every visualization, in order to reduce the chance of the model hallucinating, and maintain consistent output formats. 

We used GPT4o-mini, and queried the model using OpenAI's API using a Python script.
The scripts and the prompts can be found in the supplemental material.

\subsubsection{Results}

For the rating and each for the extracted features, we performed a t-test between the LLM-extracted values, and the manually coded features and crowdsourced rating. 
Eight features were able to be extracted by GPT4o-mini to the same level as an expert human; three features were not. 
Table~\ref{tab:gpt-feature-extraction-comparison} provides an overview of the degrees of freedom (DoF), t-statistics, and p-values from the t-tests, and highlights which features the model was unable to extract as well as an expert human. 

\begin{table}[htb]
\centering
\caption{\textbf{Overview of results from t-tests comparing human feature extraction to GPT4o-mini features extraction.}
Features that GPT4o-mini were able to extract similarly to a human are highlighted in blue.}
\label{tab:gpt-feature-extraction-comparison}
\begin{tabular}{@{}lrrr@{}}
\toprule
\textbf{Feature} & \textbf{DoF} & \textbf{t-statistic} & \textbf{p-value} \\ \midrule
\rowcolor[HTML]{DAE8FC} 
Grouped or Overlapping variant    & 1165               & -3.831      & < 0.001    \\
\rowcolor[HTML]{DAE8FC} 
Number of colors                  & 1077               & 4.134       & < 0.001 \\
\rowcolor[HTML]{DAE8FC} 
Number of data series             & 1128               & -4.216      & < 0.001 \\
3D visualization                  & 1275               & -0.452      & 0.652     \\
\rowcolor[HTML]{DAE8FC} 
Sufficient description and legend & 1264               & -13.214     & < 0.001 \\
\rowcolor[HTML]{DAE8FC} 
Single-colored background         & 1123               & 21.443      & < 0.001 \\
Multiple Coordinated Views        & 1246               & 0.693       & 0.488     \\
\rowcolor[HTML]{DAE8FC} 
Small multiples                   & 848                & 2.693       & 0.007     \\
\rowcolor[HTML]{DAE8FC} 
Multiple chart types              & 1095               & 4.358       & < 0.001 \\
Multiple scales for one axis      & 1152               & 1.540       & 0.124     \\
\rowcolor[HTML]{DAE8FC} 
Use of annotations                & 1233               & -8.719      & < 0.001 \\ \bottomrule
\end{tabular}
\end{table}

\begin{figure}
    \centering
    \includegraphics[alt={Box-and-whister plots showing the relationship between GPT4o-mini's complexity ratings, and the rations from the user study. There is a clear trend that visualizations rated complex in the study are also rated more complex by GPT4o-mini, however there is a large variety in tyhe correlation between GPT4o-mini's ratings and the ratings from the study.},width=0.9\linewidth]{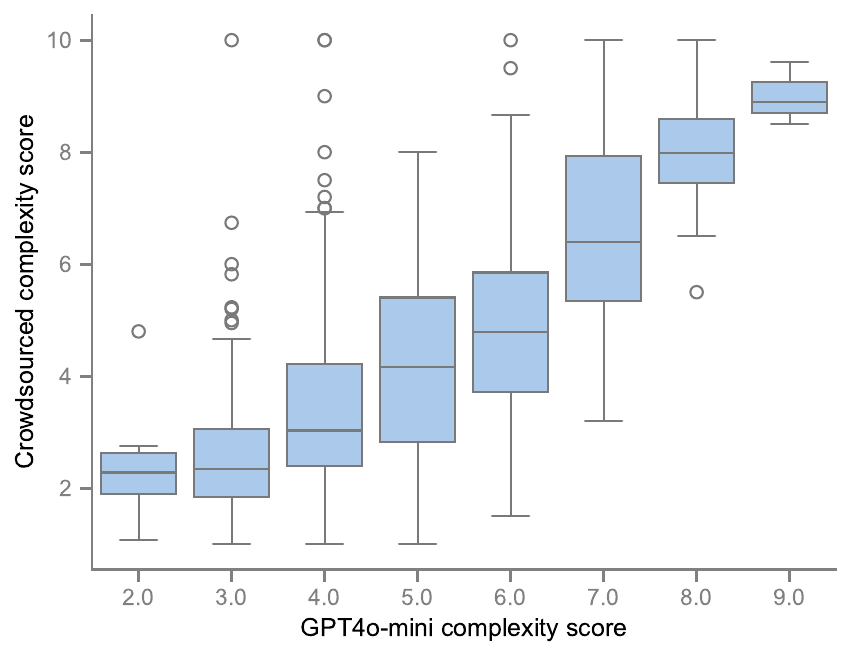}
    \caption{\textbf{Complexity score plotted against GPT4o-mini's estimated complexity score.}
    GPT4o-mini seems reasonably able to estimate the complexity of data visualizations, with some outliers.}
    \label{fig:GPT-scores-vs-complexity}
\end{figure}

A t-test between GPT4o-mini's complexity ratings and the crowdsourced complexity ratings (t(1163) = -4.277, p < 0.001) showed that the model was able to estimate the perceived visualization complexity of the crowdsourcing participants. 
Figure~\ref{fig:GPT-scores-vs-complexity} compares GPT4o-mini's ratings to the average crowdsourced ratings for each visualizations.
GPT4o-mini captures the complexity ratings reasonably well, with some variety and outliers. 
Notably, GPT4o-mini would not rate any of the visualizations as either a 1 or a 10.

We queried GPT4o-mini again to evaluate the consistency of the LLMs response. 
An intraclass correlation analysis (ICC (3,1)) reveals that the LLM is consistently able to provide a very similar rating of the visualization complexity (F(23.245) = 0.918, p < 0.01). 
We explain the results in detail, and relate them to the posed hypotheses next.

%% file: sections/06-discussion.tex
\section{Discussion}

The results of our three experiments show that the perceived complexity of visualizations can be correlated to some extend, to the features of the visualizations. 
Here we discuss our results, what they mean for our understanding of visualization complexity, and provide directions for reducing the perceived complexity. 
We also discuss whether it even makes sense to try to measure an objective complexity of visualizations, given its highly subjective nature. 
Finally we address some of the limitations present in our work.

\subsection{Explaining the Results}


Here we go through the experiments, and explain the results from each.



Experiment 1---using image analysis to estimate visualization complexity---did not yield usable results for finding a correlation between objective and subjective visualization complexity. 
As mentioned in Section~\ref{experiment1-results}, there is a weak to negligible correlation between the examined image complexity measures, and the collected visualization complexity scores. 
Here we provide our analysis of why we achieve these results. 
Since the Shannon's entropy calculation considers a histogram of the entire image, we did not expect it to yield any usable results, which was also the case. 
Mahon and Lukasiewicz's method \cite{mahon_minimum_2024}, however, uses convolutions to form clusters of similar pixels.
We expected that the well-defined structure of a visualization would be apparent in the results, but this was not the case.  
For the methods described by Rosenholtz et al.~\cite{rosenholtz_feature_2005, rosenholtz_measuring_2007} we expected that the metric would capture a lower level of visual clutter in visualizations made up of fewer primitives, such as a piechart, compared to a node-link diagram or parallel coordinate plot. 
This was not the case, with only weak correlations. 
We speculate that the individual subparts of a chart; axis, label, etc., each are interpreted as their own cluster, where the human eye might see a single cluster, due to gestalt principles. 
This would increase the amount of clusters, or clutter, in the image, and increase the image complexity scores. 
Another potential source of these results is the varied quality of the images used.  
The algorithms might pick up the noise that can be present in compressed images, which the participants did not see, and attribute it to higher image complexity. 



The analysis for Experiment 2---examining the applicability of a multilinear regression---was more successful. 
The model has a fitness of $R^2 = 0.49$, which is not particularly good, but still captures a large portion of the perceived complexities of the visualizations. 
The feature that mostly increasing the perceived complexity, according to the model, is using connection as an encoding.
This encompasses visualizations such as node-link diagrams, dendrograms, chord diagrams, arc charts, and alluvial diagrams. 
\rev{All of these visualizations, except dendrograms, have among the 12 highest complexity ratings by the participants.} 
They are likely considered complicated because they represent chart types that are relatively unfamiliar, as discussed above. 

Dendrograms were rated as much less complex, being the 15th least complex visualization, making it an outlier for this encoding. 
Perhaps the participants were familiar with this visualization type, and that it therefore was not perceived as complex. 
This illustrates how a single, or few, features are not enough for evaluating the complexity of data visualizations.
Other encodings that increased the perceived complexity was volume, position, and color saturation, which include many three-dimensional visualizations. 
As discussed earlier, adding an extra dimension to a visualization increases the complexity of the schema, which in turn can increase the viewers perceived complexity of the visualization.

Of the coded features, small multiples, multiple scales for one axis, multiple coordinated views, and multiple chart types in one visualizations, all causes an increase in a visualization's complexity. 
This is in line with hypothesis \textbf{H3}.
We argue that they constitute features that objectively increase complexity. 
All of these features increase the complexity, by adding additional encodings to the same dimensions for different elements in the schema, or by requiring the viewer to relate multiple visualizations at the same time. 
All of these features result in a composition of one or more baseline charts, which we expected to increase the complexity, as outlined in hypothesis \textit{H3}.

\begin{figure}
    \centering
    \begin{subfigure}[t]{0.45\columnwidth}
        \centering
        \includegraphics[alt={Two box-and-whister plots showing that visualizations with a legend were generally rated as less complex than visualizations without a legend.},height=6cm]{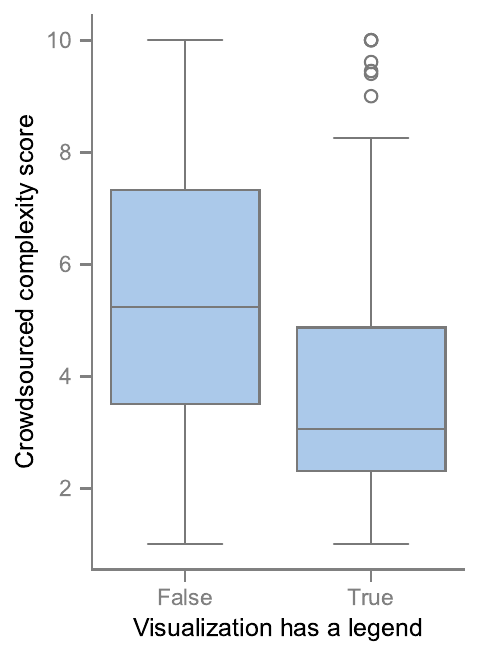}
        \caption{Complexity score for legends}
    \end{subfigure}%
    \begin{subfigure}[t]{0.45\columnwidth}
        \centering
        \includegraphics[alt={Two box-and-whister plots showing that visualizations with annotations were generally rated as more complex than visualizations without annotations.},height=6cm]{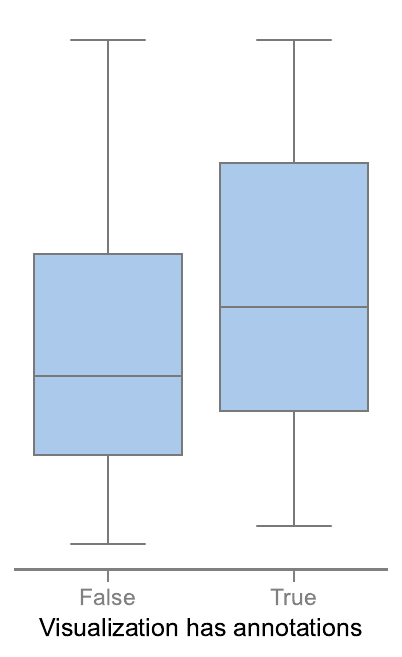}
        \caption{Complexity score for annotations}
    \end{subfigure}%
    \caption{\textbf{Complexity scores for (a) visualizations with and without legends, and (b) visualizations with and without annotations.}
    Visualizations with a legend are on average perceived as less complex than visualization without one. 
    Contrary to expectations, visualizations with annotations were perceived as more complex than those without annotations.  
    This is likely a result of these annotations being hard to understand on their own, and therefore making the visualization seem more complex to the participants.}
    \label{fig:GPT-scores-vs-annotations}
\end{figure}

We expected that a good legend as well as the use of annotations would both reduce the perceived complexity of the visualizations. 
Figure~\ref{fig:GPT-scores-vs-annotations} illustrates the distribution of complexity ratings for visualizations with and without legends, and visualizations with and without annotations. 
On average, visualizations with a legend was rated less complex than those without a legend. 
With annotations, visualizations with annotations were rated as more complex on average, than those without annotations. 
While this might suggest that annotations should be avoided, we believe that this is, in fact, a false positive. 
Many of the annotations were highly domain specific, and many annotations did not include sufficient descriptions, leaving the participant to interpret the extra information themselves. 
This illustrates that a feature considered good for comprehension, might reduce it when applied carelessly.

Experiment 3---using a vision-enabled LLM for feature extraction and complexity rating---was also mostly successful. 
As Table~\ref{tab:gpt-feature-extraction-comparison} illustrates, GPT4o-mini was able to extract most coded features from the visualizations successfully. 
The three features that were not extracted at the same performance as a human was 3D, multiple coordinated views, and use of multiple scales for one axis. 
The inability to identify 3D may be due to how the LLM interprets the prompt.
The LLM was instructed to code for 3D ``if it is a 3D chart, including 3D effects.''
But whether the model will interpret a drop-shadow or a slight perspective as 3D or not is hard to know given the non-deterministic nature of LLMs. 

Multiple coordinated views were likely unable to be extracted correctly. 
The LLM might simply have a hard time differentiating between multiple coordinated views and small multiples, which the LLM was able to correctly extract. 
Additionally, there was a low number of MCV visualizations in the dataset, meaning that a only a few incorrect codings have major effect on the results. 
With multiple scales we expect that the model might have had a hard time identifying the second scale, causing it to apply the wrong coding.
Since the visualizations are scraped from the web, there is also no single way of drawing the second axis, meaning the model might not be able to identify them all. 

Overall, GPT4o-mini was able to extract most features form the visualizations. 
In addition, the LLM was also able to rate the perceived complexity of the visualizations in a similar manner as our crowdsourced participants. 
It is noteworthy that the LLM was unable to give a score of either 1 or 10. 
This may be due to the model being coded to avoid ``extreme'' answers, preferring a more balanced response.

\subsection{Hypotheses}

Here we return to the hypotheses posed in this paper, and evaluate them based on the previous discussion of the results. 


\begin{enumerate}
    \item[\textbf{H1}] \textit{\rev{Common chart types} have low perceived complexity}
    \rev{is \textbf{partially accepted}, as most of the common chart types had low ratings.}
    
    \item[\textbf{H2}] \textit{Annotations reduces complexity} 
    is \textbf{inconclusive}, as there is likely a significant source of error in that the annotations in the dataset are often lacking any accompanying explanation. 
    
    \item[\textbf{H3}] \textit{Composition increase complexity} 
    is \textbf{accepted}, as composite features (superimposing charts and multiple coordinated views) significantly increased the perceived complexity. 
    
    \item[\textbf{H4}] \textit{Legends reduce complexity} 
    is \textbf{accepted}, as it significantly reduced the perceived complexity. 

    \item[\textbf{H5}] \textit{Visual complexity metrics correlates with perceived visualization complexity} 
    is \textbf{rejected}, as we found no correlation between the metrics and the perceived complexity. 
    
    \item[\textbf{H6}] \textit{LLMs can perform feature extraction} 
    is mostly \textbf{accepted}, as 8/11 features were successfully extracted by the LLM. 
    
    \item[\textbf{H7}] \textit{LLMs can adequately rate visualizations} 
    is \textbf{accepted}, as the LLM is statistically and consistently similar to the crowdsourced data. 
\end{enumerate}

\subsection{Visualization Complexity: Worth the Trouble?} 

With the experiments conducted in this paper, we have to ask ourselves, is it even worth it to try to measure the objective and subjective complexity of visualizations, and find correlations between the two? 
Finding a true, objective complexity score is likely not possible.
While measures such as image complexity exists, they do not take into account the purpose of visualizations; to convey information graphically to humans. 
The human perception of the visualizations, and of their complexity is intertwined with the visualization's, type, design and style, while viewer's visualization literacy, and familiarity with the chart type heavily affects how they perceive the complexity of the visualization. 
We find that there is simply too many factors that affect the complexity of a visualization, that we cannot separate from the visualization itself, without loosing sight of the goal: to make visualizations more approachable to human perception. 

Another point is that there already exists multiple guidelines for designing effective and expressive visualizations~\cite{cleveland_graphical_1984, mackinlay_automating_1986, zhu_measuring_2007, wongsuphasawat_voyager_2016}. 
These general guidelines could be considered as good enough, and that a highly effective visualization is also one that presents the data with as little complexity as possible, without losing expressiveness. 
Instead of trying to reduce the complexity of visualizations, effectively dumbing them down, the goal could instead be to find efficient ways to raise the visualization literacy of the viewer. 
Examples could include providing on-demand instructions, or onboarding the viewer for a novel visualization type in an ad-hoc manner. 

There are still good arguments for continuing the pursuit of visualization complexity. 
A better understanding of how features of a visualization affects the perceived complexity adds to our understanding of not only visualization complexity, but of how we design effective visualizations as a whole. 
Because an effective visualizations is one that reduces the cognitive load placed on the viewer, a less complex visualization schema would also be a more effective one. 
This would not only benefit novices, but also experts. 
Additionally, understanding which chart types are generally more complex than other could inform educators on how to progressively increase students' visualization literacy, by gradually increasing the complexity, and building on the newly acquired knowledge. 




\subsection{Limitations and Future Work}

This work is not without its limitations. 
Firstly, the multilinear regression model could be more robust, if more visualizations were included in the dataset.
640 visualizations spread across 50 visualizations types does not leave many visualizations per type. 
In addition, the distribution of visualization types were not even, meaning that there were many chart types with only a handful of examples, while others, like bar charts, had many examples and many different designs. 
The manual coding of a large dataset would have been prohibitively time-consuming, however.
Fortunately, as our results show, using an LLM may help us work around this issue. 

On the other hand, using an LLM for this task could be seen as bringing a gun to a knife fight; in other words, large language and foundation models are heavy-duty and blunt instruments. 
For one thing, LLMs such as GPT4o require significant computation and are thus energy-hungry. 
A more streamlined and special-purpose deep learning architecture customized to this specific task may yield less energy demands; in fact, it may even yield better performance.
For example, we could use a pipeline that processes visualization images through ResNet50 (omitting the final classification layer) while simultaneously encoding categorical features (such as chart type) and numerical features (such as number of visual elements) through embedding layers and normalization respectively, combining these feature vectors through fully connected layers to predict visualization complexity scores on a scale of 1 to 10.
Designing such a custom deep learning pipeline is left for future work.

The dataset used did not allow for tight control of the features examined in this paper. 
A custom-built dataset would allow for more unambiguous results, at the cost of some of the diversity found in real-world visualizations. 
Many features could also have been coded at the time of construction with a custom dataset. 

Additionally, the crowdsourced study has been conducted in a way that forced participants to estimate an absolute perceived complexity rating for each visualization. 
However, perceived complexity is arguably relative, and participants might have given a visualization a different rating, had they compared and ranked multiple visualizations at once. 
Having multiple participants rate each visualization was an attempt to alleviate this issue. 
\add{Finally, we did not administer a visualization literacy test, which means we cannot tell how this affects the perceived complexity of visualizations. Future studies on this is encouraged, along with studies on how other aspects of visualizations affect the perceived complexity.}


%% file: sections/07-conclusion.tex
\section{Conclusion}

We have presented a crowdsourced study for collecting perceived visualization complexity and three complementary experiments that examine techniques for objectively measuring the complexity of data visualizations. 
We also provide our definition of visualization complexity based on prior literature. 
Our results show that image analysis techniques, such as Shannon's entropy and meaningful image complexity analysis, did not have a correlation with visualization complexity.
A multilinear regression (MLR) analysis was able to estimate the complexity of the visualizations with some success. 
In addition, the vision-enabled LLM GPT4o-mini was able to correctly extract most features of a visualization as well as estimate human assigned complexity scores of visualization.
The latter means that the time-consuming task of coding a large dataset of visualizations can be avoided. 
Our MLR analysis indicates that certain encodings add more complexity than others, while features such as a description significantly reduce the perceived complexity. 
While our results indicate that some features increase complexity, and others reduce it, we cannot separate the perceived complexity from the viewer and their own visualization literacy.
With this in mind, we further discuss the pros and cons of pursuing further work into visualization complexity. 
Benefits include a deeper understanding of the cognitive impact of visualizations, which can guide the construction of easily interpretable visualizations. 
On the other hand, existing guidelines for visualization design arguably already fulfills this role to some extend. 

While this is the first step towards understanding the connection between objective and subjective complexities in visualizations, we must also acknowledge the large impact of the visualization literacy of the viewers, on their perceived complexity of the visualizations used. 
In other words, yes---visualization complexity does largely ``lie'' in the eye of the beholder.

%% file: sections/acknowledgements.tex
\acknowledgments{%
    This work was supported by Villum Investigator grant VL-54492 by Villum Fonden.
    Any opinions, findings, and conclusions expressed in this material are those of the authors and do not necessarily reflect the views of the funding agency.
}